\documentclass[aps,prb,reprint,superscriptaddress]{revtex4-1}

\usepackage{graphicx}
\usepackage{color}
\usepackage{dcolumn}   

\begin{document}



\title{
	\boldmath
Direct observation of double exchange in ferromagnetic La$_{0.7}$Sr$_{0.3}$CoO$_3$ by broadband ellipsometry
\unboldmath}


\author{P. Fri\v{s}}
\affiliation{Department of Condensed Matter Physics, Faculty of Science and Central European Institute of Technology, Masaryk University, Kotl\'a\v{r}sk\'a 2, 611 37 Brno, Czech Republic}

\author{D. Munzar}
\affiliation{Department of Condensed Matter Physics, Faculty of Science and Central European Institute of Technology, Masaryk University, Kotl\'a\v{r}sk\'a 2, 611 37 Brno, Czech Republic}

\author{O. Caha}
\affiliation{Department of Condensed Matter Physics, Faculty of Science and Central European Institute of Technology, Masaryk University, Kotl\'a\v{r}sk\'a 2, 611 37 Brno, Czech Republic}

\author{A. Dubroka}
\email[]{dubroka@physics.muni.cz}

\affiliation{Department of Condensed Matter Physics, Faculty of Science and Central European Institute of Technology, Masaryk University, Kotl\'a\v{r}sk\'a 2, 611 37 Brno, Czech Republic}


\date{\today}

\begin{abstract}
We present results of our broadband ellipsometry measurements of the optical response of ferromagnetic  La$_{0.7}$Sr$_{0.3}$CoO$_3$. Our data show that the ferromagnetic transition is accompanied by a transfer of optical spectral weight from an absorption band centered at 1.5 eV to a narrow component of the Drude-like peak. 
The associated reduction of the intraband kinetic energy is significantly larger than $k_{\rm B}T_c$, confirming that the double exchange plays a major role in the ferromagnetism of doped cobaltites.
In conjunction with results of recent theoretical studies, the temperature dependence of the Drude-like peak suggests that the double exchange is mediated by $t_{2g}$ orbitals. 
\end{abstract}
\pacs{xxx}
\maketitle
Transition metal oxides with the perovskite structure are well known for their spectacular electronic and magnetic properties such as superconductivity in cuprates and colossal magnetoresistance associated with the ferromagnetic transition in hole doped manganites~\cite{Tokura2006}. While consensus has been reached 
that the ferromagnetism in the manganites is caused by the double exchange (DE) mechanism mediated by $e_g$ electrons, the origin of ferromagnetism in related pseudocubic cobaltites La$_{1-x}$Sr$_{x}$CoO$_3$ ($0.2<x<0.5$), with comparable values of the Curie temperature ($T_{\rm C}$) reaching up to 250~K, is still being debated~\cite{Merz,SamalKumar,FuchsPRL2013,Othmen2014,Lazuta2015,Smith2016,Li2016}. The physics of the cobaltites is considerably complicated by a quasidegeneracy between several spin states of a Co ion. This is caused by a competition between the Hund's rule coupling and the crystal field splitting~\cite{Maekawa2004}. In this context, it is of high importance to find out 
whether the cobaltites exhibit optical signatures of the DE comparable to those observed in the manganites. 
Recall that the hallmark of the DE mechanism in the manganites is a lowering of the effective kinetic energy of charge carriers occurring upon the transition from the paramagnetic (PM) to the ferromagnetic (FM) state~\cite{Blundell}. It manifests itself in the optical conductivity 
as a transfer of spectral weight (SW) from a band at a finite energy in the PM state to the Drude-like peak in the FM state~\cite{Okimoto1995}. The band is due to the so-called wrong spin transition (WST)~\cite{Quijada1998,Takenaka2000,Takenaka2002} arising from hopping of carriers between sites with misaligned spins~\cite{Okimoto1995,Furukawa1995,Quijada1998,Takenaka2000,Takenaka2002,ChattopadhyayMillis2000,MichaelisMillis2003}.

In order to explore possible optical signatures of DE and contribute to the clarification of the mechanism of ferromagnetism in hole doped cobaltites, we have measured their optical response as a function of temperature ($T$) using broadband ellipsometry. The early optical study of doped cobaltites by Tokura et al.~\cite{Tokura1998} is limited to room temperature. 
We observed the SW transfer between a WST band and the Drude-like peak similar to the one reported in manganites. The fact, that the associated reduction of the intraband kinetic energy is significantly larger than $k_{\rm B}T_c$, demonstrates that ferromagnetism in cobaltites is indeed driven by a DE mechanism.
 
30~nm thin film of La$_{0.7}$Sr$_{0.3}$CoO$_3$ (LSCO) was grown by pulsed laser deposition (PLD) on 10~$\times$~10~mm$^2$ substrates (La$_{0.7}$Sr$_{0.3}$) $\times$ (Al$_{0.65}$Ta$_{0.35}$)O$_3$ (LSAT)~\cite{Alineason}. The samples were annealed $in\ situ$ at the deposition temperature of 650~$^\circ$C under 10 Torr oxygen pressure to decrease oxygen vacancy concentration~\cite{Alineason}. X-ray diffraction measurements confirmed that the film is epitaxial, see Supplemental Material (SM)~\cite{SOM}. Broadband ellipsometry measurements were performed using three ellipsometers:  Woollam VASE ellipsometer (0.6--6.5~eV),  Woollam IR-VASE ellipsometer (0.08--0.65~eV) and an in-house built ellipsometer for the far-infrared range (0.01--0.08~eV)~\cite{Ellipsometer}. For each measured photon energy value, the optical conductivity of the film was obtained from the ellipsometric angles $\Psi$ and $\Delta$ using the model of coherent interferences in a layer on a substrate. The optical response of the substrate was measured on a bare substrate. In this way, the optical response was obtained without extrapolations needed for a Kramers-Kronig analysis used, e.g. in reflectivity data analysis. The film thickness was determined using ellipsometry and X-ray diffraction. The magnetic moments were measured using a vibrating sample magnetometer.

\begin{figure}[t]
	\centering
	\vspace*{-0.1cm}
	\hspace*{-0.3cm}
	\includegraphics[width=8.6cm]{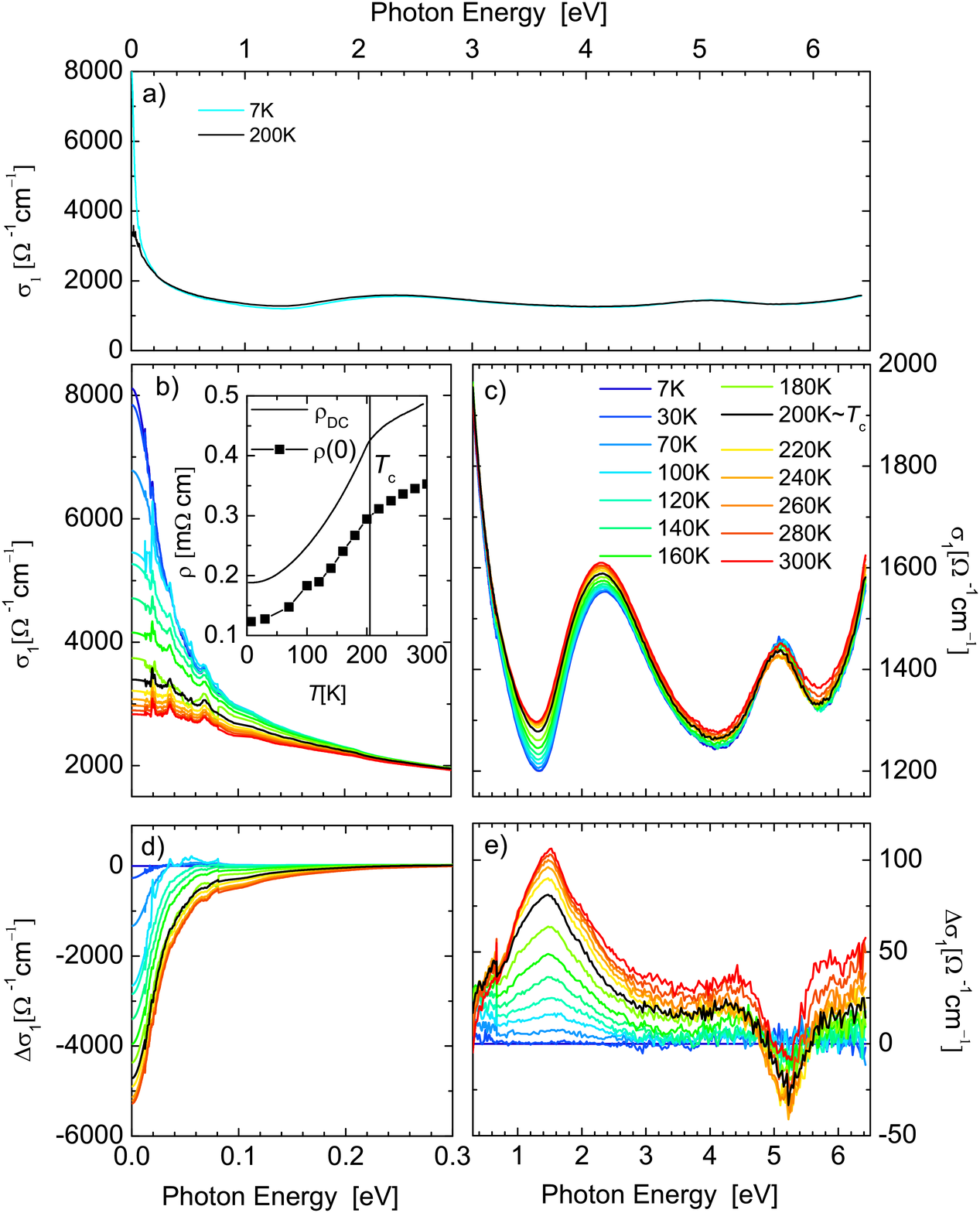}
	\vspace*{-0.2cm}
	\caption{
		The real part of the optical conductivity, $\sigma_1(\omega)$, of La$_{0.7}$Sr$_{0.3}$CoO$_3$ deposited on LSAT.  a) $\sigma_1(\omega)$ at $T=200~{\rm K}\approx T_c$ and at 7~K. b), c) $\sigma_1(\omega)$ for all measured temperatures on a magnified scale. The inset of b) shows the $T$ dependencies of $\rho(0)=1/\sigma_1(0)$ and of the DC resistivity $\rho_{\rm DC}$.
		Panels d) and e) display $\Delta \sigma_{1}(\omega,T)\equiv \sigma_{1}(\omega,T)-\sigma_{1}(\omega,7~{\rm K})$.}
	\label{fig:Sigma1odd}
\end{figure}
The resulting spectra of the real part of the optical conductivity, $\sigma_1$, of LSCO deposited on LSAT are shown in Fig.~\ref{fig:Sigma1odd}a) for $T=200$~K close to $T_{\rm C}=205$~K and for the lowest $T$ of our measurements (7~K). Figures~\ref{fig:Sigma1odd}b) and ~\ref{fig:Sigma1odd}c) show all measured spectra on a magnified scale.  The values of $\sigma_1(\omega)$ for energies lower than 0.01~eV were obtained using extrapolations based on a Kramers-Kronig consistent model introduced below. The corresponding spectra of the PLD target were found to be very similar~\cite{SOM}. The spectra of $\sigma_1$ exhibit a Drude-like peak at low energies and two absorption bands centred around 2.3~eV and 5.1~eV, see Fig.~\ref{fig:Sigma1odd}a). Based on the results of Ref.~\cite{Jeong-scientificreport}, we assign the absorption band centred at 2.3~eV to the O~$2p$ to Co~$3d$ charge transfer transition. At low $T$, $\sigma_1$  exhibits metallic behaviour with fairly high values of $\sigma_1(0)$ reaching $\sim8000\ \Omega^{-1}$cm$^{-1}$ for $T\rightarrow0$ [see Fig.~\ref{fig:Sigma1odd}(b)], but even above $T_{\rm c}\sim 200$~K, the low frequency response is still very metallic with high values of $\sigma_1(0)$ of $\sim3000\ \Omega^{-1}$cm$^{-1}$. The spectra do not exhibit any  signatures of Jahn-Teller localization (for details, see SM~\cite{SOM}), in agreement with results of Ref.~\cite{Sundaram2009}. The inset of Fig.~\ref{fig:Sigma1odd}(b) displays the $T$ dependencies of $\rho(0)=1/\sigma_1(0)$ and the directly measured DC resistivity $\rho_{\rm DC}$. The values of $\rho(0)$ agree within 30\% with the resistivity measured on a single crystal~\cite{Aarbogh2006}, which demonstrates that our thin film is of high quality and indicates that its optical properties are fairly close to the bulk ones. The magnitude of $\rho_{\rm DC}$ is by a factor of about 1.4 larger than that of $\rho(0)$ which is an effect typically observed on thin films due to linear defects~\cite{Quijada1998}. The $T$ dependencies of $\rho_{\rm DC}$ and $\rho(0)$ display a clear anomaly at $T_c\sim 205$~K similar to that of the single crystal~\cite{Aarbogh2006}, indicating that double exchange should be considered as a plausible mechanism for the magnetic ordering.

Next, we address in detail the $T$ dependence of  $\sigma_1$. 
Figures~\ref{fig:Sigma1odd}d) and~\ref{fig:Sigma1odd}e) show differential spectra: $\Delta \sigma_{1}(\omega,T)\equiv \sigma_{1}(\omega,T)-\sigma_{1}(\omega,7~{\rm K})$.
With increasing temperature, $\Delta \sigma_1$ strongly decreases below 0.1~eV and, on the contrary, a band is formed with the maximum at ca 1.5~eV. This is analogous to the DE related SW redistribution occuring in manganites, where the FM$\rightarrow$PM transition is associated with a SW transfer from the Drude peak to a band centered at 3.2~eV~\cite{Quijada1998, Takenaka2000}.
 
\begin{figure}[t]
	\centering
	\hspace*{-0.5cm}
	\includegraphics[width=10cm]{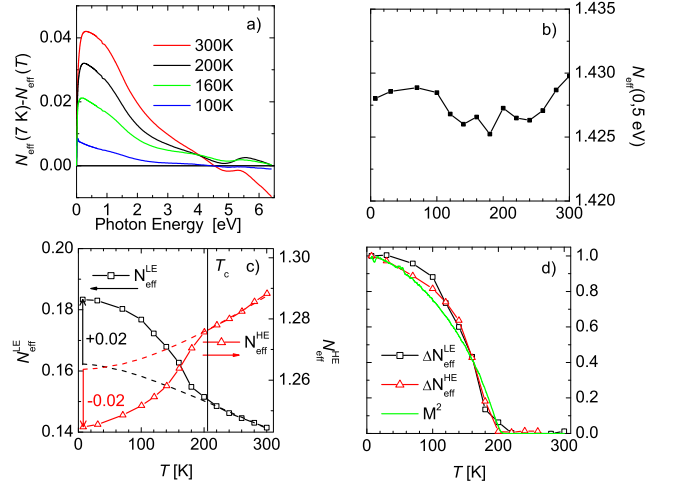}
	\caption{a) Energy dependence of the quantity $N{_{\rm eff}} (0,\omega,7~{\rm K})-N{_{\rm eff}}(0,\omega,T)$ introduced in the text.
	b) Temperature dependence of the spectral weight $N_{\rm eff}(0, 5~{\rm eV}, T)$. 
	c) $T$ dependences of the low and high energy spectral weights $N_{\rm eff}^{\rm LE}=N_{\rm eff}(0, 0.3~{\rm eV})$ and $N_{\rm eff}^{\rm HE}= N_{\rm eff}(0.3, 5~{\rm eV})$. Estimates for the background $T$ dependences discussed in the text are shown as the dashed lines. 
	d) Absolute values of the differences $\Delta N_{\rm eff}^{\rm LE}$ and $\Delta N_{\rm eff}^{\rm HE}$ between the spectral weights and the corresponding background $T$ dependences together with the square of the magnetization measured at $B=20$~mT. All quantities are normalized to their lowest temperature values.}
	\label{fig:NeffInt}
\end{figure}
We quantified the frequency and $T$ dependence of the optical SW in terms of the effective number of charge carriers per unit cell, $N_{\rm eff}$, defined as
\begin{equation}
\label{specweight}
N_{\rm eff}({\omega_L},{\omega_H})=\frac{2m_0V}{\pi e^2}\int_{\omega_L}^{\omega_H} \sigma_1(\omega){\rm d} \omega,
\end{equation}
where $m_0$ is the electron mass, $V$ is unit cell volume and $e$ is the electron charge. The quantity $N_{\rm eff}({\omega_L},{\omega_H})$ represents an estimate for the number of charge carriers per unit cell responsible for absorption between $\omega_L$ and $\omega_H$. 
Figure~\ref{fig:NeffInt}a) displays the energy dependence of the difference 
$N{_{\rm eff}} (0,\omega, T=7~{\rm K})-N{_{\rm eff}} (0,\omega, T)$
demonstrating that the SW between 0 and $\hbar\omega_H=5$~eV is essentially temperature independent.  Figure~\ref{fig:NeffInt}b) confirms that the $T$~dependence of $N{_{\rm eff}} (0,5\ {\rm eV})$ is very weak. Figures~\ref{fig:Sigma1odd}b) and~\ref{fig:Sigma1odd}c) show that the data exhibit an isosbestic point (a point where $\sigma_1$ is almost $T$ independent) at $\hbar\omega$=0.3~eV. Motivated by this finding, we plot in Fig.~\ref{fig:NeffInt}c) the $T$ dependence of the low energy SW, 
$N_{\rm eff}^{\rm LE}$, defined as $N_{\rm eff}(0,\ 0.3$~eV) and that of the high energy SW, $N_{\rm eff}^{\rm HE}$, defined as $N_{\rm eff}(0.3~{\rm eV},\ 5~{\rm eV})$. Both quantities display a pronounced anomaly at $T\approx$~200~K~$\approx T_C$. This finding demonstrates that a part of the SW transfer between high and low energies is due to the FM transition and implies that the onset of the FM order is associated with a kinetic energy saving characteristic of the DE mechanism.

The $T$ dependences of the SWs in Fig.~\ref{fig:NeffInt}c) are, in addition to the DE, influenced by a common narrowing of the Drude peak due to the increase of the quasi-particle lifetime with decreasing $T$. We attempted to separate the two contributions by approximating the ``normal" components of the $T$ dependences with extrapolations based on fits of the high temperature segments using a background function~\cite{SOM}. In Fig.~\ref{fig:NeffInt}c) they are represented by the dashed lines. The estimated magnitude of the DE related contribution, $N^{\rm DE}_{\rm eff}$ [see the vertical arrows in  Fig.~\ref{fig:NeffInt}c)], is approximately  $0.02$. 
Figure~\ref{fig:NeffInt}d)  displays the $T$ dependences of $\Delta N_{\rm eff}^{\rm LE}$, $\Delta N_{\rm eff}^{\rm HE}$ (the SWs with the background contributions subtracted) together with the $T$ dependence of $M^2$, where $M$ is the measured magnetisation. The $T$ dependences are very similar which indicates that the observed FM related SW changes are connected to the energetics of the FM transition~\cite{Blundell}.

\begin{figure}[t]
	\centering
	\includegraphics[width=8.6cm]{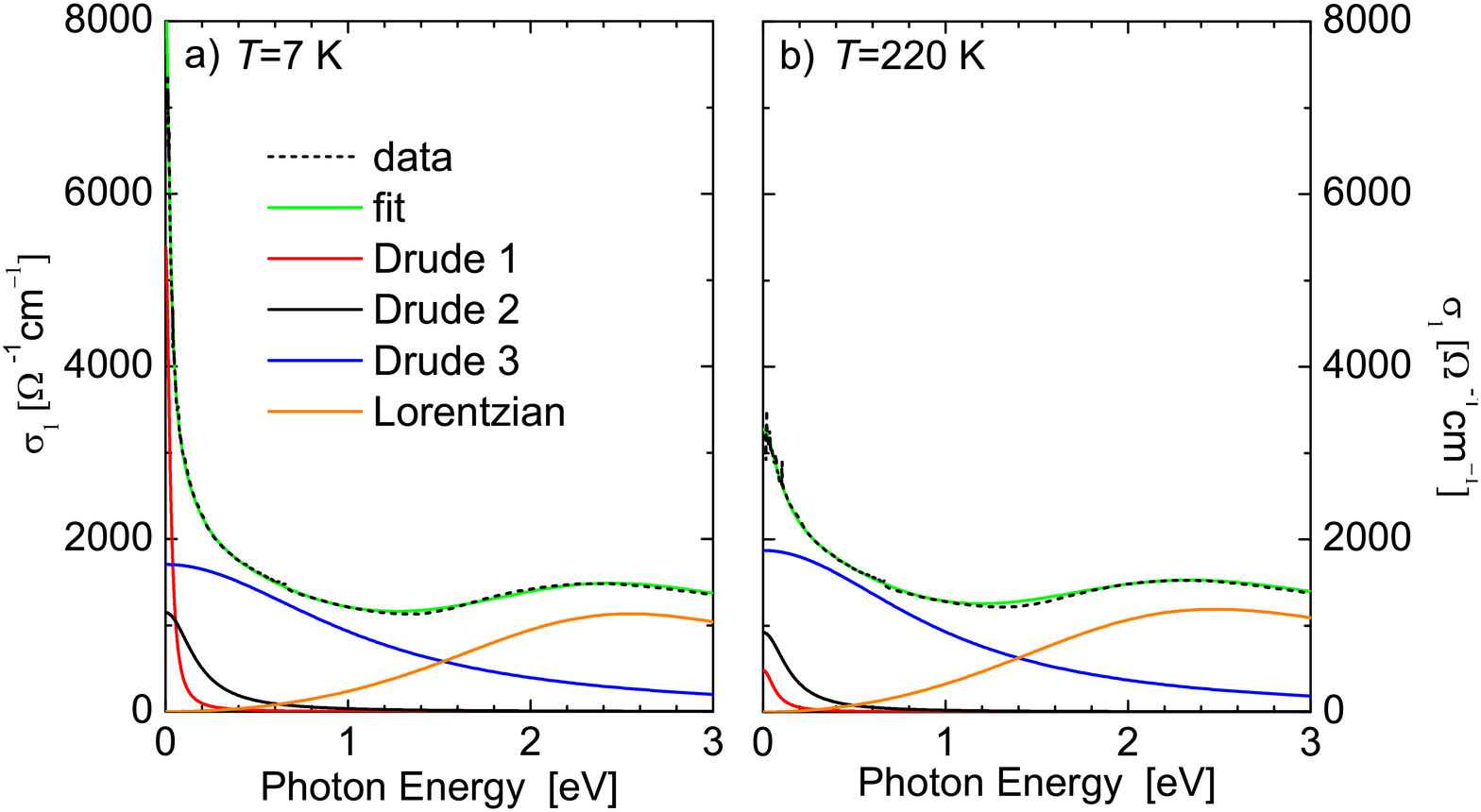}
	\caption{Measured $\sigma_1(\omega)$ (dotted line), model spectrum (solid green line) and contributions of individual terms in Eq.~(\ref{diel}) for  $T=7\ {\rm K}$ (a) and for $T=220\ {\rm K}$ (b).}
	\label{fig:Oscilators}
\end{figure}

In order to obtain additional insight into the optical response, we modeled the complex conductivity $\sigma(\omega)=-{\rm i}\epsilon_0\omega(\epsilon(\omega)-1)$ using the Drude-Lorentz model of the dielectric function 
\begin{equation}
\label{diel}
\epsilon(\omega)=1-\sum_j \frac{\omega_{D,j}^2}{\omega(\omega+{\rm{i}} \gamma_{D,j})}+\sum_k\frac{\omega_{L,k}^2}{\omega_{0,k}^2-\omega^2-{\rm i} \omega \gamma_{L,k}},
\end{equation}
where the second and the third terms on the right side represent the Drude and the Lorentz contributions, respectively, parameters $\omega_{D,j}$ and $\omega_{L,k}$ are plasma frequencies, $\gamma_{D,j}$ and $\gamma_{L,k}$ broadening parameters, and $\omega_{0,k}$ frequencies of the Lorentz terms. Figures~\ref{fig:Oscilators}a) and~\ref{fig:Oscilators}b) show the experimental data and fits for $T=7~$K  and $220\ {\rm K}> T_{\rm c}$ on a magnified scale. The complete set of obtained parameter values can be found in SM~\cite{SOM}. It turns out that the interband transitions at 2.3~eV and 5.1~eV can be represented by only one Lorentz term per transition. However, in order to obtain a reasonable fit of the free carrier response, three Drude terms are required with $\hbar\gamma_{D}=0.03$, 0.17 and 1~eV at 7~K (0.07, 0.15 and 1~eV at 220~K) and with the corresponding SWs proportional to $(\hbar\omega_{D})^2=1.1$, 1.5 and 14~eV$^2$ at 7~K (0.25, 1.0 and 14~eV$^2$ at 220~K), respectively. 
It appears that only the two narrower Drude terms change significantly with $T$  whereas the broadest Drude term is almost $T$ independent.

This behaviour can be qualitatively understood based on results of a recent LDA+DMFT study~\cite{Augustinsky}, which indicates that both $e_g$ and $t_{2g}$ bands cross the chemical potential. The $e_g$ band has a large bandwidth and its quasiparticles are strongly damped by correlations and the corresponding broadening of the spectral function is about 0.6~eV. In contrast, the $t_{2g}$ band has a relatively small bandwidth and its quasiparticles are much less damped, with the broadening of about 0.04~eV.
In view of these results we suggest that the broadest Drude term mainly corresponds to the response of the $e_g$ bands and the two narrower Drude terms correspond to the response of the $t_{2g}$ bands. The SW of the broadest Drude term is about 5 times higher than the sum of the SWs of the two narrower Drude terms, which correlates with the large difference between the predicted (e.g. in Ref.~\cite{Augustinsky}) bandwidths of the two channels.

Note that in manganites at low temperature, the Drude peak is also composed of a narrow and a broad component~\cite{Takenaka2002}. However, the components exhibit a similar and pronounced temperature dependence, which suggests that they originate from the same band. 
Relatively, the sum of the SWs of the two narrower Drude terms increases by a factor of two when going from $T_c$ to the lowest temperature, which is comparable to the two-fold increase of the intraband SW occuring in manganites~\cite{Quijada1998}.
The magnitude of the total SW transfer upon the FM transition, $N^{\rm DE}_{\rm eff}=0.02$, is approximately 7 times lower than the one observed in manganites~\cite{Quijada1998,Takenaka2000,Takenaka2002}, for details see SM~\cite{SOM}. This difference is presumably due to the narrow bandwidth of the $t_{2g}$ bands compared to the $e_g$ bands in manganites. 
The fact, that below $T_c$ spectral weight $N^{\rm DE}_{\rm eff}$ is transferred only to the narrow component of the Drude peak, is consistent with the DE  mediated by $t_{2g}$ electrons, as recently suggested~\cite{Merz,FuchsPRL2013}. The associated saving of the intraband kinetic energy, $\Delta K=3\hbar^2 N^{\rm DE}_{\rm eff}/a_0^2m_0$~\cite{Basov2005,SOM}, where $a_0=3.82$~\AA\ is the pseudocubic lattice parameter, is about 30~meV. Albeit smaller than in manganites, this value is larger than $k_{\rm B}T_{\rm C}\approx17$~meV, which shows that the kinetic energy reduction plays an important role in the mechanism of ferromagnetism in doped cobaltites.

\begin{figure}[t]
	\centering
	\includegraphics[width=7.6cm]{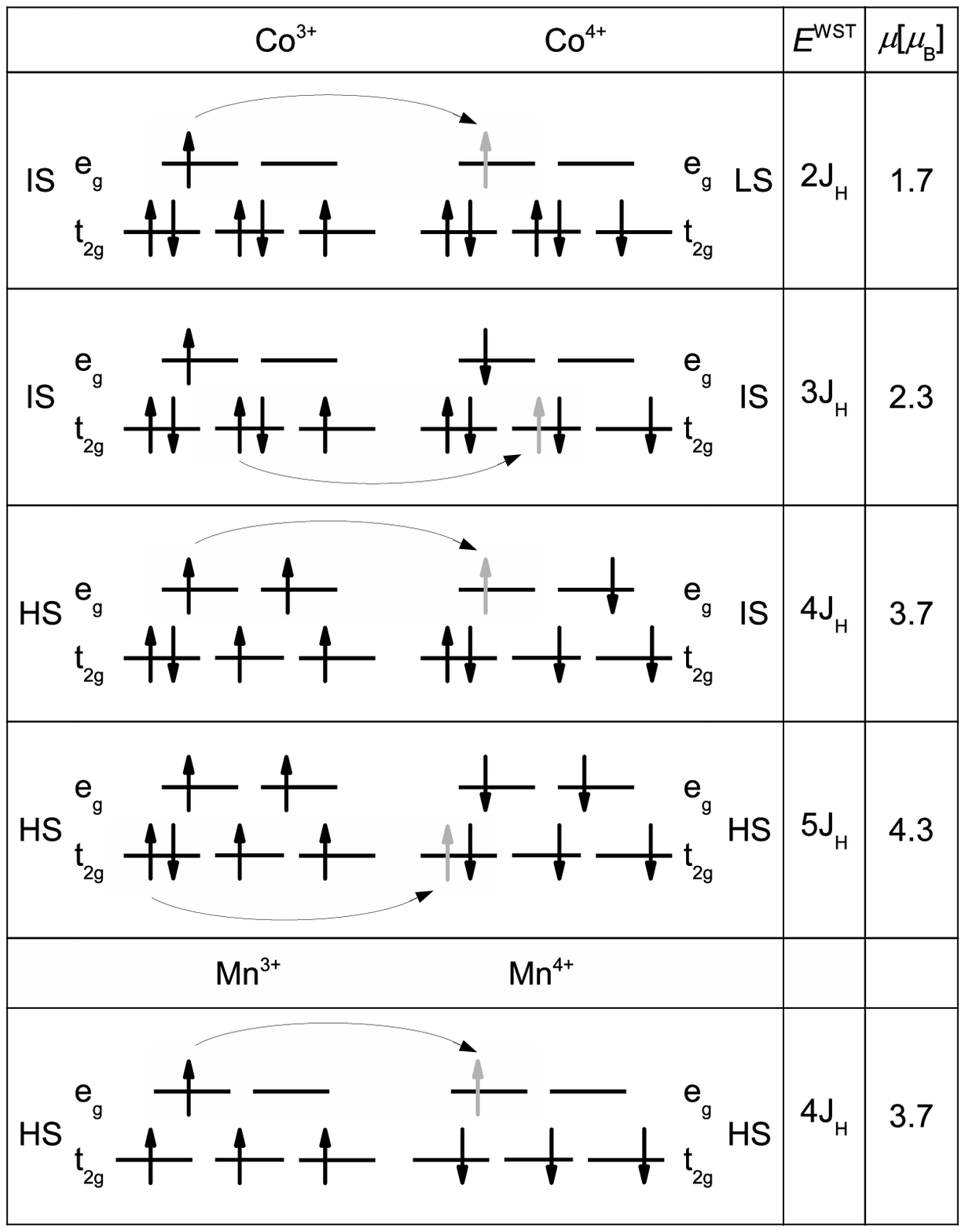}
	\caption{Examples of “wrong spin” transitions, in the ionic limit, between a Co$^{3+}$ site and Co$^{4+}$ site for various Co spin states. 
	Transition energy values, $E^{\rm WST}$, and the average magnetic moments $\mu$ per site computed for 30\% hole doping. The lowest row shows the corresponding situation in La$_{0.7}$Sr$_{0.3}$MnO$_3$.}
	\label{fig:skoky}
\end{figure}
Next, we address the origin and the properties of the 1.5 eV band in Fig.~\ref{fig:Sigma1odd}e).
As discussed above, the $T$ dependence of its SW correlates with that of the narrow Drude peak, in a way remarkably similar to that of the WST feature in manganites.
Based on this observation we suggest that the 1.5 eV band is also a result of a WST. To clarify the nature of this WST, we first consider, for the sake of simplicity,  weakly coupled pairs of Co sites with well defined spin states~\cite{Lee2005Comentary}. Examples of corresponding WST from a Co$^{3+}$ ion  to a  Co$^{4+}$ ion are shown in Fig.~\ref{fig:skoky}.
The abbreviations IS and HS in the Co$^{3+}$ column [LS, IS, HS in the Co$^{4+}$ column]
stand for the intermediate ($t_{2g}^5$ $e_g^1$) and high spin ($t_{2g}^4$ $e_g^2$) configurations of Co$^{3+}$ [low ($t_{2g}^5$ $e_g^0$), intermediate ($t_{2g}^4$ $e_g^1$) and high spin ($t_{2g}^3$ $e_g^2$) configurations of Co$^{4+}$].
The WSTs, indicated by the long arrows, result in excited states with one unpaired electron whose spin is oriented antiparallel to those of other unpaired electrons at the same site. Note that the optical process preserves the electron spin. For comparison, the bottom panel of Fig.~\ref{fig:skoky} shows the WST between Mn$^{3+}$ and Mn$^{4+}$ occurring in manganites. Finally, the energies 
$E^{\rm WST}$ of the WST and the average values of the magnetic moment $\mu$ per transition metal site are given in the second and the third columns, respectively.  
The energies have been obtained using the Hund's rule coupling Hamiltonian~\cite{Hamiltonian}
\begin{equation}
H=-J_{\rm H} \sum_{l,l', l \not=l'} {\bf S}_{l}{\bf S}_{l'},
\end{equation}
where the sum runs over all pairs $l,l'$ of 3d orbitals, $J_{\rm H}$ is the coupling constant and ${\bf S}_{l}$ is the spin operator for the orbital $l$. 
Clearly, the WST appear due to the presence of pairs of transition metal sites with misaligned spins.  
In the FM phase, the complete alignment of spins allows for the coherent transport of charge carriers. The WST is absent and the spectral weight it had in the PM state is transferred to the Drude peak. The corresponding lowering of the effective kinetic energy is the essence of the DE mechanism. Note that the value of $E^{\rm WST}$ can be reduced with respect to that expected within the ionic limit due to a partially itinerant character of charge carriers. For manganites, this has been suggested to reduce the value of $E^{\rm WST}$ from $4J_{\rm H}$ to ca $2J_{\rm H}$~\cite{Furukawa1995,ChattopadhyayMillis2000,MichaelisMillis2003,LinMillis2008}.

Importantly, in Fig.~\ref{fig:skoky}  $E^{\rm WST}$ is approximately proportional to $\mu$. Provided the values of $J_{\rm H}$(Co) and $J_{\rm H}$(Mn) are approximately the same, the values of $E^{\rm WST}/\mu$ for LSCO and LCMO can be expected to be similar.
Indeed, for LSCO the experimental values, $E^{\rm WST}_{\rm LSCO}=1.5$~eV and $\mu\approx1.5\ \mu_{\rm B}$~\cite{SamalKumar,SOM},  yield 
$E^{\rm WST}/\mu=1.0$~eV/$\mu_{\rm B}$, which is comparable to the value 
of $0.8$~eV/$\mu_{\rm B}$ for LSMO. The latter has been obtained using $E^{\rm WST}_{\rm LSMO}=3.2$~eV~\cite{Quijada1998} and $\mu_{\rm LSMO}\approx3.6$~ $\mu_{\rm B}$~\cite{Martin1996}.  
In the most simple picture, the facts that $E^{\rm WST}_{\rm LSCO}$ is about one half of 
$E^{\rm WST}_{\rm LSMO}$ and that $\mu_{\rm LSCO}\approx1.5\ \mu_{\rm B}$ would point to the first row of Fig.~\ref{fig:skoky}.
Results of a recent X-ray study~\cite{Merz}, however, indicate that the Co ions are partially in the LS and partially in the HS configuration, rather than in the IS one. This point of view is supported by the LDA+DMFT study~\cite{Augustinsky}, which argues that the average occupations of HS-related and LS-related states are approximately 30\% and 70\%, respectively~\cite{Augustinsky}. 
This offers the following qualitative picture of the DE in cobaltites: the DE is connected to the HS states and meditated by the $t_{2g}$ electrons. As a consequence of the admixture of low spin states, $E^{\rm WST}$ can be expected to be considerably lower than $5J_{\rm H}$ of the extreme HS-HS case (shown in the fourth row in Fig.~\ref{fig:skoky}).

In summary, we have observed that the ferromagnetic transition in La$_{0.7}$Sr$_{0.3}$CoO$_3$ is associated with a transfer of spectral weight from an absorption band centered at 1.5 eV to a narrow component of the Drude peak. 
Similarly to the manganites, the band can be interpreted in terms of a wrong spin transition involving Co sites with misaligned spins. Its energy is lower than that of the corresponding band in manganites which is consistent with a lower value of the ordered magnetic moment. The fact that the FM influences only the narrow and relatively weak component of the Drude peak, in conjunction with results of a recent theoretical study~\cite{Augustinsky}, suggests that the double exchange is mediated by $t_{2g}$ orbitals. The associated reduction of the intraband kinetic energy is significantly larger than $k_{\rm B}T_c$, confirming that the double exchange is indeed at the heart of ferromagnetism in doped cobaltites.\\
\begin{acknowledgments}
We acknowledge helpful discussions  with J. Chaloupka, D. Fuchs, G. Khaliullin and K. Kn\'\i\v{z}ek and magnetic measurements at IPM ASCR supervised by M. Hapla. This work was financially supported by the MEYS of the Czech Republic under the project  CEITEC 2020 (LQ1601) and carried out with the support of CEITEC Nano Research Infrastructure (MEYS CR, 2016--2019).
\end{acknowledgments}

\renewcommand\thefigure{S\arabic{figure}}   

\newpage

\begin{widetext}
	\center
	\large
	{\bfseries
		\boldmath
		Supplemental material for ``Direct observation of double exchange in ferromagnetic La$_{0.7}$Sr$_{0.3}$CoO$_3$ by broadband ellipsometry"
		\unboldmath}
	\vspace{0.5cm}
\end{widetext}
\
\boldmath
\subsection{X-ray diffraction data on La$_{0.7}$Sr$_{0.3}$CoO$_3$ thin film}
\unboldmath

\begin{figure}[b]
	a)\includegraphics[width=7.5cm]{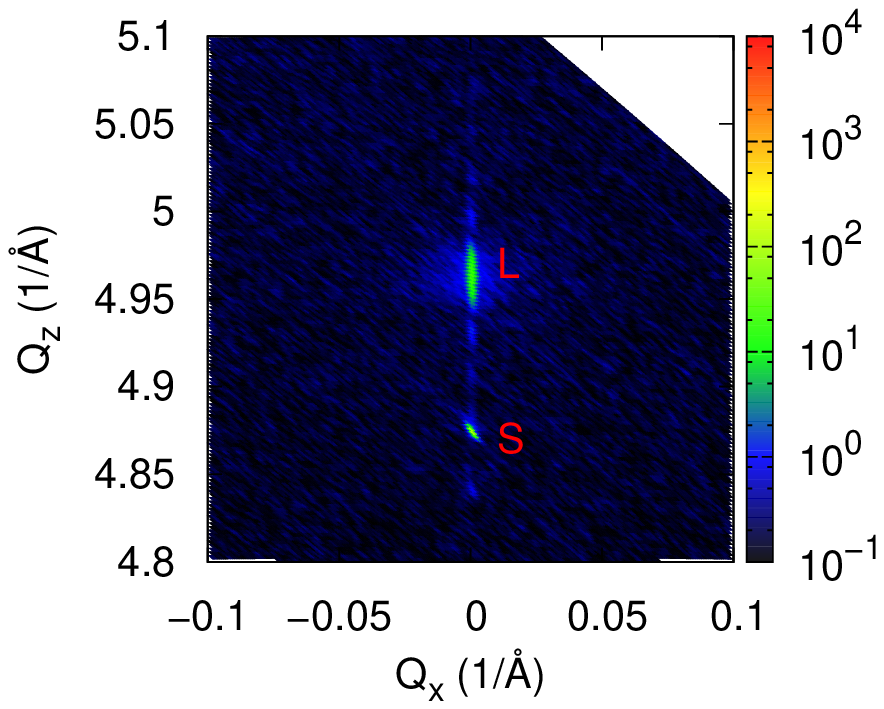}
	b)\includegraphics[width=7.5cm]{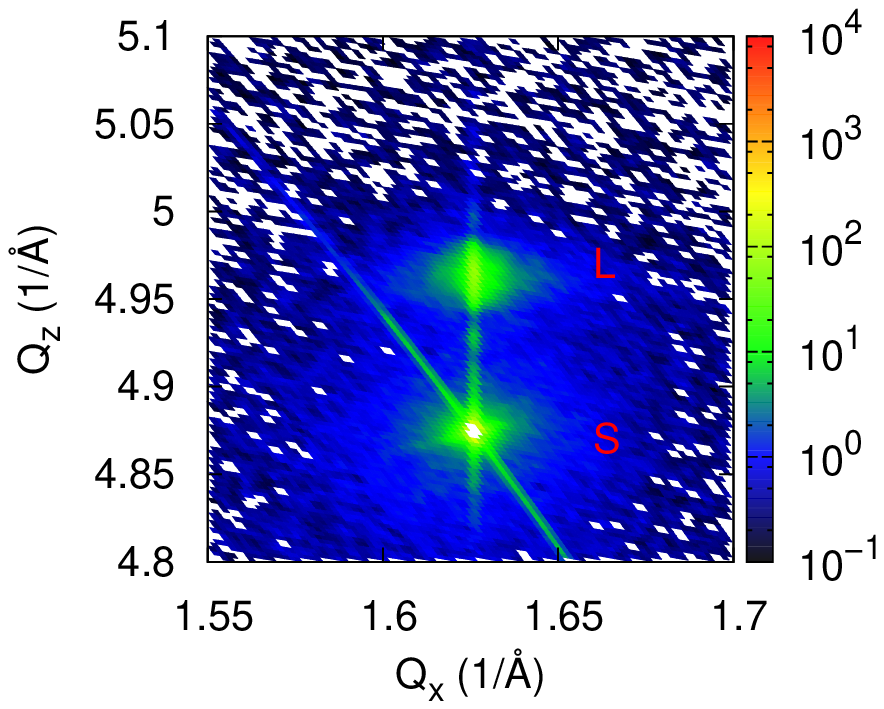}
	\caption{Reciprocal space maps of the La$_{0.7}$Sr$_{0.3}$CoO$_3$ thin film in the vicinity of 003 (a) and 103 (b) reciprocal lattice points, respectively. LSAT substrate and La$_{0.7}$Sr$_{0.3}$CoO$_3$ epitaxial layer peaks are denoted by "S" and "L" marks, respectively.
	}
	\label{sfig1}
\end{figure}

The structure of the La$_{0.7}$Sr$_{0.3}$CoO$_3$ thin film deposited on LSAT substrate has been characterized using high resolution x-ray diffraction.
We have used Rigaku SmartLab diffractometer equipped with a copper x-ray tube, parabolic multilayer mirror, 2-bounce Ge(220) monochromator and Dtex solid state detector.
The resulting reciprocal space maps measured in the vicinity of 003 and 103  reciprocal lattice points at room temperature, are shown in Fig.~\ref{sfig1}a) and b) respectively.
The layer peak, L, and the substrate peak, S, are in the same $Q_x$ position in both reciprocal space maps indicating perfect pseudomorphic growth of the layer.
The value of the lattice parameter of the substrate has been determined as $a_\mathrm{LSAT}=(3.8665\pm0.0002)$\,\AA{} and that of the out-of-plane lattice parameter of the layer is $a_{\perp}=(3.7975\pm0.0005)$\,\AA{}.
Using the value of the pseudocubic lattice parameter of polycrystalline LSCO of $a_\mathrm{0}=3.822$\,\AA{}, we have obtained the in-plane strain component $\epsilon_{xx}=1.2\,\%$ and the out-of-plane strain component $\epsilon_{zz}=-0.6\,\%$. 

\begin{figure}
	\includegraphics[width=7.5cm]{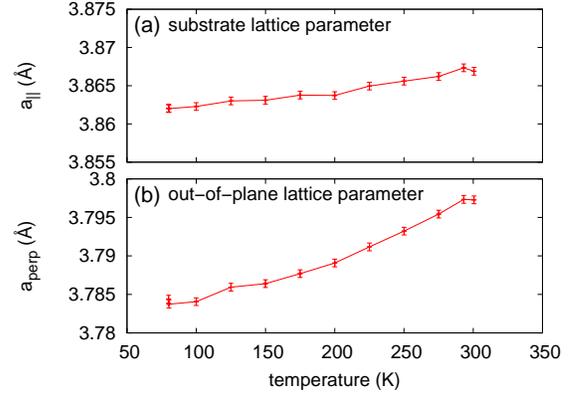}
	\caption{Temperature dependences of the LSAT substrate lattice parameter (a) and of the out-of-plane lattice parameter of the La$_{0.7}$Sr$_{0.3}$CoO$_3$ thin film (b). The in-plane lattice parameter of the La$_{0.7}$Sr$_{0.3}$CoO$_3$ epitaxial thin film equals to the LSAT substrate lattice parameter as shown in reciprocal space maps in Fig.~\ref{sfig1}.
	}
	\label{slattice}
\end{figure}

\begin{figure}
	\includegraphics[width=7.5cm]{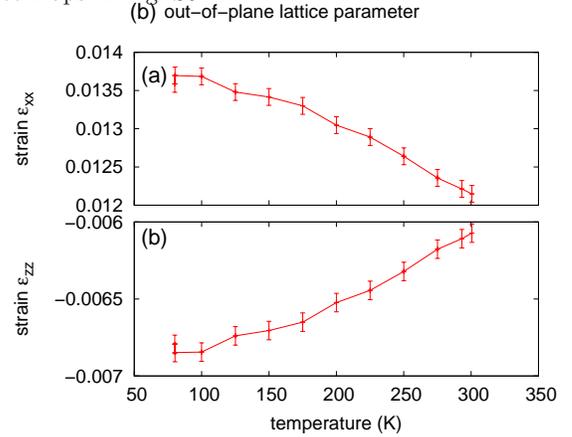}
	\caption{Temperature dependences of the in-plane (a) and out-of-plane (b) strain components of the La$_{0.7}$Sr$_{0.3}$CoO$_3$ thin film.
	}
	\label{sstrain}
\end{figure}

Furthermore, we have measured the positions of 002, 003 and 103 x-ray diffraction peaks as a function of temperature in order to determine the temperature dependence of the strain. 
The measurements were performed using Anton Paar TTK450 liquid nitrogen cooled x-ray cryostat. The resulting temperature dependences of the substrate lattice parameter and of the out-of-plane thin film lattice parameter are shown in Fig.~\ref{slattice}a) and \ref{slattice}b), respectively. Since the thin film is laterally lattice matched to the LSAT substrate, their in-plane lattice parameters are equal.
Using the Poisson ratio value of 0.25, we have determined the temperature dependences of the thin film in-plain and out-of-plain strain components
$\epsilon_{xx}$ and $\epsilon_{zz}$ shown in Fig.~\ref{sstrain}a) and~\ref{sstrain}b), respectively. The thermal expansion coefficient of the layer is higher than the substrate one which leads to about 10\% 
increase of the elastic strain with cooling from 300~K to 77~K.

\subsection{Discussion of Jahn-Teller localisation}
At low temperatures, $\sigma_1$ of the thin film exhibits metallic behaviour with fairly high values of $\sigma_1(\omega=0)$ reaching $\sim8000\ \Omega^{-1}$cm$^{-1}$ for $T\rightarrow0$~K,
see Fig.~1b) of the main text. This value is comparable to the value of approximately 9000~$\Omega^{-1}$cm$^{-1}$ reported for La$_{0.7}$Sr$_{0.3}$MnO$_3$ (LSMO)~\cite{Okimoto1997,Quijada1998}. Above $T_{\rm C}\sim200$~K, the low frequency response is still very metallic with high values of $\sigma_{\rm DC}$ of $\sim3000~\Omega^{-1}$cm$^{-1}$. These values are significantly higher than the corresponding value for LSMO at $T_C$ of approximately 300~$\Omega^{-1}$cm$^{-1}$, see Ref.~\cite{Okimoto1997}. The low conductivity in the PM state of manganites is caused by a partial charge localization due to the Jahn-Teller (JT) distortion which also produces a strong band at approximately 1~eV~\cite{Quijada1998, Takenaka2002}. Since our PM state spectra are very metallic and do not exhibit any pronounced band below 1~eV, we conclude that in LSCO the charge localization due to the JT effect is absent or not significant, which is in agreement with results obtained using local structure techniques~\cite{Sundaram2009}.

\subsection{Saturated magnetic moment}

\begin{figure}
	\centering
	\includegraphics[width=8.6cm]{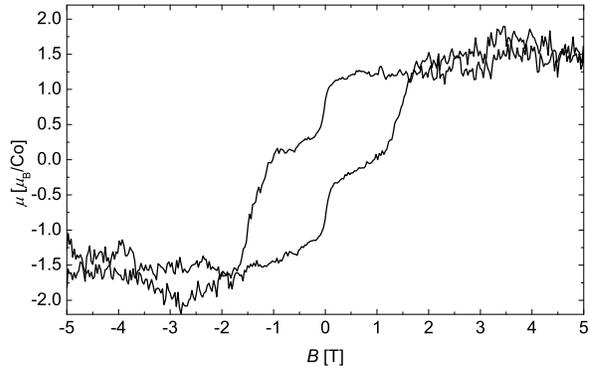}
	\caption{
		Magnetic moment per Co site of the La$_{0.7}$Sr$_{0.3}$CoO$_3$ thin film at 4~K, as a function of the applied field, obtained using vibrating sample magnetometer. 
	}
	\label{Magnetizace}
\end{figure}
Figure~\ref{Magnetizace} displays the magnetic moment per Co site of the La$_{0.7}$Sr$_{0.3}$CoO$_3$ thin film at 4~K obtained using vibrating sample magnetometer as a function of the field $B$ applied  along the sample surface. A diamagnetic response from the LSAT substrate was subtracted. The saturated magnetic moment is about 1.5~$\mu_{\rm B}$, which is a typical value for bulk La$_{0.7}$Sr$_{0.3}$CoO$_3$~\cite{SamalKumar}. This indicates that the magnetic state of our thin film is close to that of the magnetic state of bulk La$_{0.7}$Sr$_{0.3}$CoO$_3$. The hysteresis loop exhibits two steps that are likely due to differences in the coercive fields~\cite{PrietoRuiz2015} between the material close to the substrate--thin film boundary and close to the surface.

\subsection{Spectral weight considerations}
The ``normal" components of the $T$ dependent spectral weights shown in Fig.~2c) of the main text and in Fig.~\ref{targetMag} have been obtained by extrapolations based on fits of the segments above $T_c$, using the background function $f(T)=a+b\sqrt{1+(T/T_0)^2}$. This function allows one to approximate both the high temperature quasi-linear trend and the quadratic dependence occurring at low temperatures, expected in case of an electron-electron scattering.

Several values of $N^{\rm DE}_{\rm eff}$ in manganites can be found in the literature: 0.22 for La$_{0.7}$Sr$_{0.3}$MnO$_3$  (LSMO) thin film (the value has been determined from the data at 2.7~eV in Fig.~3 of Ref.~\cite{Quijada1998}), 0.145 for La$_{0.825}$Sr$_{0.175}$MnO$_3$ bulk (the value has been determined as the sum of $S_{2-4}$ and $S_{4-7}$ of Fig. 3b) of Ref.~\cite{Takenaka2000}) and  0.13 for La$_{0.6}$Sr$_{0.4}$MnO$_3$ bulk (the value determined from the data at 2.7~eV in the inset of Fig.~3 in Ref.~\cite{Takenaka2002}). Note, however, that these estimates 
have not been corrected for the ``normal" narrowing of the Drude-like peak.
Based on our analysis of the data of the LSCO film contained in the main text, we expect that also in the manganites, the normal contribution to 
$N^{\rm DE}_{\rm eff}$ is not larger than 30\%. It implies that the corrected FM related SW change $N^{\rm DE}_{\rm eff}$ in manganites is in the range from ca 0.1 to ca 0.2, where the scatter is due to the scatter in the literature data. This gives $5\leq N^{\rm DE}_{\rm eff}({\rm LSMO})/N^{\rm DE}_{\rm eff}({\rm LSCO})\leq 10$. Alternatively, one could compare the values before the correction procedure, which yields a very similar outcome.

Concerning the effective kinetic energy: note that  Refs.~\cite{Quijada1998,LinMillis2008} contain discussions of the kinetic energy along a particular direction, that is given by (see, e.g. Ref.~\cite{Basov2005})
\begin{equation}
	K_x=-\frac{2\hbar^2a_0}{\pi e^2}\int_{0}^{\omega_H} \sigma_1(\omega){\rm d} \omega=-\hbar^2 N_{\rm eff}/a_0^2m_0\ .
\end{equation} 
The total kinetic energy of an isotropic material $K$ is three times higher, i.e., 
\begin{equation}
	K=-3\hbar^2 N_{\rm eff}/a_0^2m_0\ .
\end{equation}

\subsection{Values of the fitting parameters obtained by fitting the thin film data presented in the main text}

\begin{figure}
	\centering
	\includegraphics[width=8.6cm]{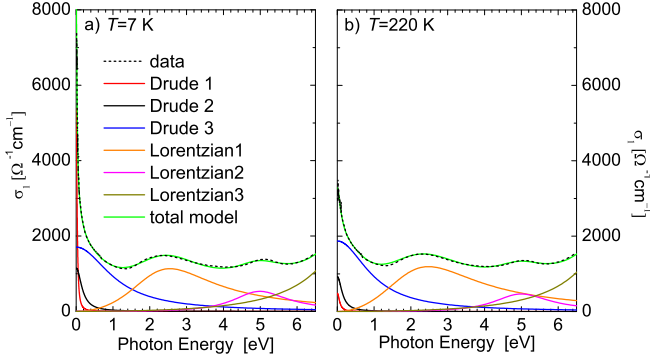}
	\caption{
		Measured $\sigma_1(\omega)$ (dotted line), model spectrum (solid green line) and contributions of individual terms in Eq.~(2) of the main text at $T=7\ {\rm K}$ (a) and at $T=220\ {\rm K}$ (b).}
	\label{FitFull}
\end{figure}

\begin{table}[b]
	\caption{\label{tab:table2}Parameter values (in eV) as determined by the best fit of the data using the model specified in the main text. The error bars represent two standard deviations.}
	\begin{ruledtabular}
		\begin{tabular}{ccc}
			Drude terms & 7~K & 220~K \\
			\hline
			$ \omega_{D,1}$ & 1.04 $\pm$ 0.01 &  0.5 $\pm$ 0.5 \\
			$ \gamma_{D,1}$ & 0.027 $\pm$ 0.001  & 0.07 $\pm$ 0.02 \\
			$ \omega_{D,2}$ & 1.23 $\pm$ 0.02 & 1.0 $\pm$ 0.1 \\ 
			$\gamma_{D,2}$ & 0.177 $\pm$ 0.006 & 0.15 $\pm$ 0.03 \\ $\omega_{D,3}$ & 3.73 $\pm$ 0.04 & 3.72 $\pm$ 0.02 \\
			$\gamma_{D,3}$ & 1.09 $\pm$ 0.03 & 0.99 $\pm$ 0.02 \\
			\hline
			Lorenz terms  & 7~K & 220~K\\
			\hline
			$\omega_{L,1}$ & 4.9 $\pm$ 0.4 & 5.3 $\pm$ 0.3\\
			$\omega_{0,1}$ & 2.55 $\pm$ 0.04 & 2.48 $\pm$ 0.04\\
			$\gamma_{L,1}$ & 2.8 $\pm$ 0.2 & 3.2 $\pm$ 0.2\\
			$\omega_{L,2}$ & 2.7 $\pm$ 0.5 & 2.5 $\pm$ 0.3\\
			$\omega_{0,2}$ & 4.99 $\pm$ 0.07 & 5.00 $\pm$ 0.05\\
			$\gamma_{L,2}$ & 1.8 $\pm$ 0.4 & 1.8 $\pm$ 0.2\\
			$\omega_{L,3}$ & 9.93 $\pm$ 0.08 &  9.92 $\pm$ 0.08\\
			$\omega_{0,3}$ & 8.66  & 8.66  \\
			$\gamma_{L,3}$ & 2.6 $\pm$ 0.2 & 2.5 $\pm$ 0.2\\
		\end{tabular}
	\end{ruledtabular}
\end{table}

For completeness, we display in Fig.~\ref{FitFull} the thin film data, the fit and the individual terms of Eq.~(2) of the main text for the whole measured frequency range. The full set of the obtained parameter values  is presented in Table~\ref{tab:table2}.

\boldmath
\subsection{Data from polycrystalline La$_{0.7}$Sr$_{0.3}$CoO$_3$ PLD target}
\unboldmath
\begin{figure}
	\centering
	\includegraphics[width=8.6cm]{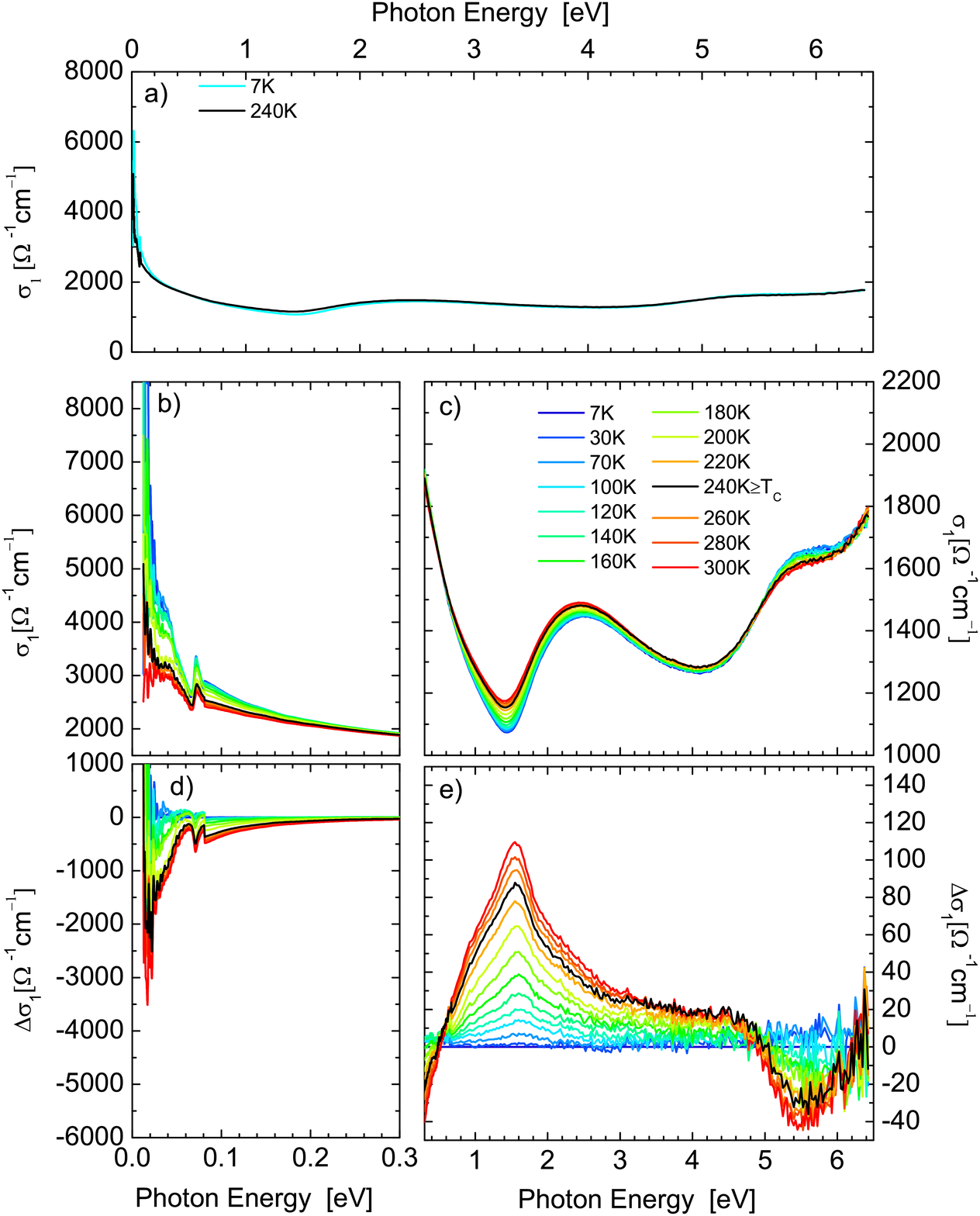}
	\caption{
		The real part of the optical conductivity, $\sigma_1(\omega)$, of La$_{0.7}$Sr$_{0.3}$CoO$_3$ polycrystalline PLD target.  a) $\sigma_1(\omega)$ at $T=240~{\rm K}\geq T_c$ and at 7~K. b), c) $\sigma_1(\omega)$ for all measured temperatures on a magnified scale. 		Panels d) and e) show $\Delta \sigma_{1}(\omega,T)\equiv \sigma_{1}(\omega,T)-\sigma_{1}(\omega,7~{\rm K})$.
	}
	\label{target}
\end{figure}

\begin{figure}
	\centering
	\includegraphics[width=8.6cm]{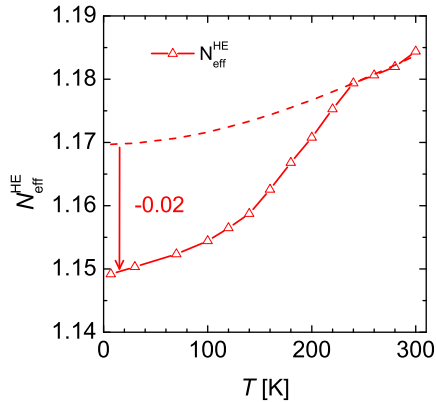}
	\caption{High energy spectral weight, $N_{\rm eff}^{\rm HE}= N_{\rm eff}(0.5, 5~{\rm eV})$, of the polycrystalline PLD target as a function of temperature. An estimate for the background function is shown as the dashed line. }
	\label{targetMag}
\end{figure}
Figure~\ref{target} shows optical data of the polished polycrystalline La$_{0.7}$Sr$_{0.3}$CoO$_3$ PLD target with $T_c=220$~K used for the deposition. Overall, the data are very similar to those of the thin film presented in the main text. There are minor differences, particularly in the far-infrared range, where the phonon near 0.07 eV has an anomalous line shape and an enhanced SW because of the effective medium effects due to grain boundaries. Nevertheless, the WST band  visible in the differential spectra in Fig.~\ref{target}e) is very similar to that of the thin film: the band maximum occurs at the same energy of about 1.5~eV and the temperature dependence of the SW between 0.5~eV (i.e. the energy of the isosbestic point) and 5~eV shows the same 0.02 difference with respect to the background [see Fig.~\ref{targetMag}]. 
This suggests that the epitaxial strain, induced by the relatively small lattice mismatch between LSAT and the thin film  (1.2\%), does not significantly alter magnetic and electronic properties of doped La$_{0.7}$Sr$_{0.3}$CoO$_3$, in agreement with the observation of Ref.~\cite{FuchsPRL2013}.

\bibliographystyle{apsrev4-1}
\bibliography{bibliography}
\nocite{Okimoto1997}
\nocite{PrietoRuiz2015}
\end{document}